\documentclass{article}

\input{tcilatex}
\begin{document}

\title{Comment on \textquotedblleft Bell Experiment with Genuine Energy-Time
Entanglement\textquotedblright\ by Cabello \textit{et al}.: Why Franson's
Experiment Violates Local Realism}
\author{Luiz Carlos Ryff \\
\textit{Instituto de F\'{\i}sica, Universidade Federal do Rio de Janeiro, }\\
\textit{Caixa Postal 68528, 21941-972 RJ, Brazil}\\
{\small e-mail: ryff@if.ufrj.br}}
\maketitle

\begin{abstract}
It is shown that the model introduced by Cabello\textit{\ et al}. to
criticize the Franson experiment suffers from the same weakness as the
previous model introduced by Aerts \textit{et al}. with the same purpose. It
is also shown why we can assume with confidence that the Franson experiment
does violate local realism.
\end{abstract}

In a recent article \textrm{[1]}, Cabello \textit{et al}. introduced a local
model which violates a Bell-CHSH inequality \textrm{[2]} for the Franson
experiment \textrm{[3]}. However, their model presents the same deficiency
as the model proposed by Aerts \textit{et al}. \textrm{[4]}. As has been
pointed out \textrm{[5]}, to see this we can consider a Franson experiment
in which the long path L$_{1}$ (followed by photon $\nu _{1}$) and the long
path L$_{2}$ (followed by photon $\nu _{2}$) have different lengths, $L_{1}$
and $L_{2}$, respectively. Now, even though the four possible combinations
of photon paths, S$_{1}$-S$_{2}$, S$_{1}$-L$_{2}$, L$_{1}$-S$_{2}$, and L$%
_{1}$-L$_{2}$ are distinguishable, the two above models still predict
violations of a Bell-CHSH inequality. In consequence, they are totally
unsatisfactory (there seems to be no reason to perform an experiment to see
this). This also makes evident that \textquotedblleft to conceive a local
realistic model for Franson's experiment is far from being an easy task, 
\textit{even} considering real inefficient detectors, since the model has to
predict an interference when $L_{1}=L_{2}$ \textit{and} no interference when 
$L_{1}\neq L_{2}$\textquotedblright \textrm{[5]}. As a result, a model
similar to the one discussed by Clauser and Horne \textrm{[6] }and that
could mimic the quantum mechanical predictions\textrm{\ }for the Franson
experiment seems much more difficult to be devised.

Franson's experiment can also be modified in order to make evident some of
the nonlocal aspects of entangled photons (e.g., the conflict between
wave-like and nonlocal properties\textrm{[7]}, and the possibility of
obtaining interaction-free which-path information \textrm{[8]}). For that
reason, a local model also has to mimic the quantum mechanical predictions
for these experiments, that is, it has to explain why the two-photon
interference is no longer observed. This makes evident the versatility of
the Franson experiment. Local models should display the same versatility.

\end{document}